\documentclass[aps,prd,twocolumn,groupedaddress,floatfix,showpacs]{revtex4}

\usepackage{graphicx}

\begin{document}

%%\preprint{}

\title{Gravitational lensing effects on the gamma-ray burst Hubble diagram}

\author{Masamune Oguri$^{1}$ and Keitaro Takahashi$^{2}$}
%\email[]{oguri@astro.princeton.edu}
%\thanks{}
\affiliation{
$^1$Department of Astrophysical Sciences, Princeton University, Princeton, NJ 08544. \\
$^2$Department of Physics, Princeton University, Princeton,  NJ 08544.
}

\date{\today}

\begin{abstract}
Gamma-Ray Bursts (GRBs) offer a potential way to extend the
Hubble diagram to very high redshifts and to constrain the nature of
dark energy in a way complementary to distant type Ia supernovae. 
However, gravitational lensing systematically brightens distant GRBs
through the magnification bias, in addition to increasing the
dispersions of distance measurements. We investigate how the
magnification bias limits the cosmological usage of GRBs. 
We perform Monte-Carlo simulations of {\it Swift} GRBs assuming a
cosmological constant dominated universe and then constrain the dark
energy equation of state neglecting gravitational lens effects. 
The originally assumed model is recovered with 68\% confidence limit 
even when the dispersion of inferred luminosities is comparable to
that of type Ia supernovae. This implies that the bias is not so
drastic for {\it Swift} GRBs as to change constraints on dark energy
and its evolution. However, the precise degree of the bias in
cosmological parameter determinations depends strongly on the shape of
the luminosity function of GRBs. Therefore, an accurate determination
of the shape of the luminosity function is required to remove the
effect of gravitational lensing  and to obtain an unbiased Hubble
diagram. 
\end{abstract}

%\keywords{}

\maketitle

%%%%%%%%%%%%%%%%%%%%%%%%%%%%%%%%%%%%%%%%%%%%%%%%%%%%%%%%%%%
%%%%%%%%%%%%%%%%%%%%%%%%%%%%%%%%%%%%%%%%%%%%%%%%%%%%%%%%%%%
\section{Introduction}
%%%%%%%%%%%%%%%%%%%%%%%%%%%%%%%%%%%%%%%%%%%%%%%%%%%%%%%%%%%
%%%%%%%%%%%%%%%%%%%%%%%%%%%%%%%%%%%%%%%%%%%%%%%%%%%%%%%%%%%

The expansion history of the universe provides invaluable information
about the energy budget. Observations of distant type Ia supernovae
(SNeIa) revealed that the universe is accelerating \cite{Riess:1998cb},
implying that the significant amount of dark energy is present.
The nature of dark energy is still unknown and is one of central issues
in modern cosmology. It is expected that detailed measurements of the
expansion history leads to understanding the nature of dark
energy, because the expansion rate of the universe is related with the
equation of state of the dark energy. 

In addition to SNeIa, gamma-ray bursts (GRBs) can serve as another
probe of the expansion history of the universe. Although luminosities
of different long duration GRBs are not quite similar, they can be
inferred from observables such as variability \cite{Fenimore:2000vs},
spectral-lag \cite{Norris:1999ni}, peak energy
\cite{Amati:2002ny,Yonetoku:2003gi}, and jet opening angle
\citep{Frail:2001qp}. An advantage of GRBs is that GRBs are observed
at higher redshifts than SNeIa; the average redshift of GRBs
discovered by {\it Swift} satellite is 
$z\sim 3$ \cite{Jakobsson:2005jc}, and GRB with redshift as high as
$z=6.3$ was indeed discovered \cite{Haislip:2005mr}. The higher mean
redshift implies that we may obtain useful information on the expansion
history from observations of GRBs, in a complementary way to SNeIa
\cite{Takahashi:2003ap}. In fact, attempts to construct the Hubble
diagram from observed GRBs and to constrain cosmological parameters have
already been made \cite{Schaefer:2002sx}. Given the fact
that the number of GRBs is now rapidly growing, GRBs are expected to
offer unique insight into the nature of dark energy.  

However, it should be kept in mind that gravitational lensing has
sometimes a great impact on high-redshift objects. While the effect of
gravitational lensing on SNeIa is modest, effectively just introducing
additional dispersion (e.g., \cite{Holz:2004xx}), gravitational lensing
is expected to affect GRBs much more drastically because of the
following two reasons.  First, the effect of lensing is a strong
function of the redshift. At higher redshifts, the probability
distribution functions (PDFs) of lensing magnification has much larger
dispersions, and also deviates from the Gaussian distribution more
significantly. Second, and more importantly, their apparent isotropic
luminosity could range over three orders of magnitudes, although they
may be standardized through the use of luminosity indicators to some
(uncertain) level. In this situation, the magnification bias, i.e.,
the effect of observing highly magnified GRBs selectively whose fluxes
without lensing effect are below a flux limit of GRB detections,
becomes quite significant.  

In this paper, we study the effect of gravitational lensing on the GRB
Hubble diagram and the determination of the dark energy equation of
state. We pay particular attention on the non-Gaussian nature of
magnification PDFs and the magnification bias. First, we derive the
amount of magnification bias analytically, and then we perform
Monte-Carlo simulations to show the impact of gravitational lensing on
the cosmological parameter determination.  As a fiducial
cosmological model, we consider a model with the matter density
$\Omega_{\rm M}=0.27$, the dark energy density $\Omega_{\rm DE}=0.73$,
the dark energy equation of state $w(z)=-1$, the dimensionless Hubble
constant $h=0.72$.  Below we refer the model as the $\Lambda$CDM. 

%%%%%%%%%%%%%%%%%%%%%%%%%%%%%%%%%%%%%%%%%%%%%%%%%%%%%%%%%%%
%%%%%%%%%%%%%%%%%%%%%%%%%%%%%%%%%%%%%%%%%%%%%%%%%%%%%%%%%%%
\section{Magnification Probability Distribution}
%%%%%%%%%%%%%%%%%%%%%%%%%%%%%%%%%%%%%%%%%%%%%%%%%%%%%%%%%%%
%%%%%%%%%%%%%%%%%%%%%%%%%%%%%%%%%%%%%%%%%%%%%%%%%%%%%%%%%%%

\begin{figure}[!t]
\includegraphics[width=0.48\textwidth]{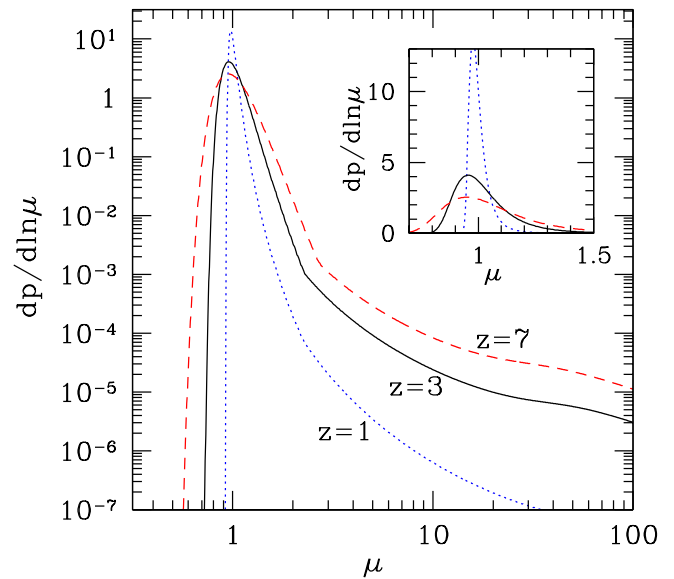}
\caption{The magnification PDFs for source redshift $z=1$ ({\it
 dotted}), $3$  ({\it solid}), and  $7$  ({\it dashed}) are plotted as a
 function of the magnification factor $\mu$. The inset shows expanded
 view of the PDFs around the peaks.} 
\label{fig:pdf}
\end{figure}

In this section, we derive the magnification PDF which is used to
examine the effect of gravitational lensing on GRBs. We adopt a
compound method that we derive the PDFs at around the peak and tail
separately and combine them. First, for the PDF at the peak, we
construct it from the PDF of lensing convergence $\kappa$. We adopt 
a modified log-normal model \cite{Das:2005yb} for the convergence PDF: 
\begin{widetext}
\begin{equation}
P(\kappa)d\kappa=\frac{N}{\sqrt{2\pi}\omega}
\exp\left[-\frac{\{\ln(1+\kappa/|\kappa_{\rm min}|)+\omega^2/2\}^2
\left(1+A/(1+\kappa/|\kappa_{\rm min})\right)}{2\omega^2}\right]
\frac{d\kappa}{\kappa+|\kappa_{\rm min}|},
\end{equation}
\end{widetext}
where $\kappa_{\rm min}$ is the convergence of the empty beam, and $N$,
$A$, $\omega$ are determined from the normalization and the conditions
$\int\kappa P(\kappa)d\kappa=0$ and
$\int\kappa^2P(\kappa)d\kappa=\langle \kappa^2\rangle$.  Although the
model was intended to describe the PDF for each lens plane, we regard
it as the PDF for convergence projected along all line-of-sight by
adopting the projected variance for $\langle \kappa^2\rangle$; Taruya
et al. \cite{Taruya:2002vy} showed that this prescription well
reproduces the non-Gaussianity of the PDF at around the peak,
$\kappa/\langle\kappa^2\rangle^{1/2}<10$. The projected variance of
lensing convergence, $\langle\kappa^2\rangle$ is computed by a
standard method using non-linear power spectrum. Since GRBs are point
sources, we adopt sufficiently small smoothing angle 
$\theta_s=0.1''$ in computing the variance. We convert it to the
magnification PDF by neglecting shear and adopt a relation
\begin{equation}
\mu=\frac{1}{(1-\kappa)^2},
\end{equation}
where $\mu$ denotes a magnification factor. On the other hand, we
compute PDFs at high-magnification tails with a halo approach assuming
an NFW profile for dark halos (see, e.g.,
\cite{Perrotta:2001zm}). Specifically, we compute the cross section,
i.e., the area with the magnification larger than $\mu$,
$\sigma(>\mu)$, assuming an NFW profile \cite{Navarro:1996he}, and
then they are summed up with a mass function of dark halos $dn/dM$:
\begin{equation}
P(\mu)d\mu=-\frac{d}{d\mu}\left[\int dz_l \int dM
  \sigma(>\mu)\frac{dl}{dz_l} \frac{dn}{dM}\right].
\end{equation}
For the mass function, we adopt a form proposed by Sheth \& Tormen
\cite{Sheth:1999mn}. These PDFs are computed assuming the $\Lambda$CDM
model and $\sigma_8=0.8$. We connect both PDFs where they intersect, which
occurs at $\mu\sim 2$. We show the resulting PDFs in Figure
\ref{fig:pdf}. As seen, the widths of the PDFs are wider for higher
redshifts. The probabilities for relatively high magnifications are
becoming very high at high redshifts. The peaks shift to smaller
magnification factor $\mu$ with increasing redshift to assure that the
mean magnification is unity. These behaviors clearly indicate that effects of
gravitational lensing is more prominent at high redshifts. The PDFs are
roughly consistent with those in literature (e.g.,
\cite{Holz:2004xx}). We note that the detailed shape of the
magnification PDF is not very important for our purpose; our qualitative
result will not be affected by a slight change of the PDF. 

%%%%%%%%%%%%%%%%%%%%%%%%%%%%%%%%%%%%%%%%%%%%%%%%%%%%%%%%%%%%%%%%%%%%%
%%%%%%%%%%%%%%%%%%%%%%%%%%%%%%%%%%%%%%%%%%%%%%%%%%%%%%%%%%%%%%%%%%%%%

%%%%%%%%%%%%%%%%%%%%%%%%%%%%%%%%%%%%%%%%%%%%%%%%%%%%%%%%%%%
%%%%%%%%%%%%%%%%%%%%%%%%%%%%%%%%%%%%%%%%%%%%%%%%%%%%%%%%%%%
\section{Magnification Bias}
%%%%%%%%%%%%%%%%%%%%%%%%%%%%%%%%%%%%%%%%%%%%%%%%%%%%%%%%%%%
%%%%%%%%%%%%%%%%%%%%%%%%%%%%%%%%%%%%%%%%%%%%%%%%%%%%%%%%%%%

In this section, we estimate the impact of the magnification bias effect
analytically. This can be done by combining the magnification PDF we
derived above and GRB luminosity function in the following way. 
The GRB luminosity function $\phi(L)$ has been constrained from the
number count of GRBs as a function of fluxes and the redshift
distribution of GRBs
\cite{Yonetoku:2003gi,Schmidt:1999iw,Firmani:2004fn}. We adopt a
luminosity function derived by Firmami et al. \cite{Firmani:2004fn}
which includes the luminosity evolution of GRBs. Specifically we adopt
a model of a double power-law luminosity function with no correlation
of peak energy with luminosity (a model named 'DPE' in
\cite{Firmani:2004fn}). The slopes of the luminosity function are
$-1.53$ and $-3.4$, and the break luminosity $L_{\rm b}$ evolves with
redshift such that $L_{\rm b}\propto (1+z)^{1.2}$. They also
constrained the GRB formation rate from the data, and pointed out that
it resembles the observed cosmic star formation rate.

Given the shape of luminosity function, we can estimate the effect of
the magnification bias. The {\it mean} magnification factor for
observed GRBs are calculated from the magnification PDF as 
%%%%%%%%%%%%%%%%%%%%%%
\begin{equation}
\langle\mu\rangle=\frac{\int \mu (dp/d\mu)f(\mu)d\mu}{\int (dp/d\mu)f(\mu)d\mu},
\end{equation}
%%%%%%%%%%%%%%%%%%%%%%
%%%%%%%%%%%%%%%%%%%%%%
\begin{equation}
f(\mu)=\frac{\int_{L_{\rm min}}^\infty
    \phi(L/\mu)dL/\mu}{\int_{L_{\rm min}}^\infty\phi(L)dL},
\end{equation}
%%%%%%%%%%%%%%%%%%%%%%
where $L_{\rm min}$ is the minimum luminosity corresponding to the
flux limit of observations and is a function of the redshift. 
As a specific example, we consider the {\it Swift} satellite as a GRB
detector. Since all GRBs with the redshifts measured have observed
photon flux larger than $\sim 0.4$ ${\rm cm^{-2}s^{-1}}$ ($15-150$ keV) 
\cite{swift}, in what follows we adopt the flux limit $P_{\rm lim}=0.4$
${\rm cm^{-2}s^{-1}}$. The energy band is converted correctly using a
spectral energy distribution of GRBs presented in
\cite{Firmani:2004fn}. The magnification factor due to lensing affects
the estimate of a distance modulus by $-2.5\log\mu$. Thus, we expect
that the Hubble diagram derived from GRBs is on average shifted from
the true one by $-2.5\log\langle\mu\rangle$. Hence this quantity
provides a good estimate of how much constraints on cosmological
parameters are biased. 

%%%%%%%%%%%%%%%%%%%%%%%%%%%%%%%%%%%%%%%%%%%%%%%%%%%%%%%%%%%%%%%%%%%%%
\begin{figure}[!t]
\includegraphics[width=0.48\textwidth]{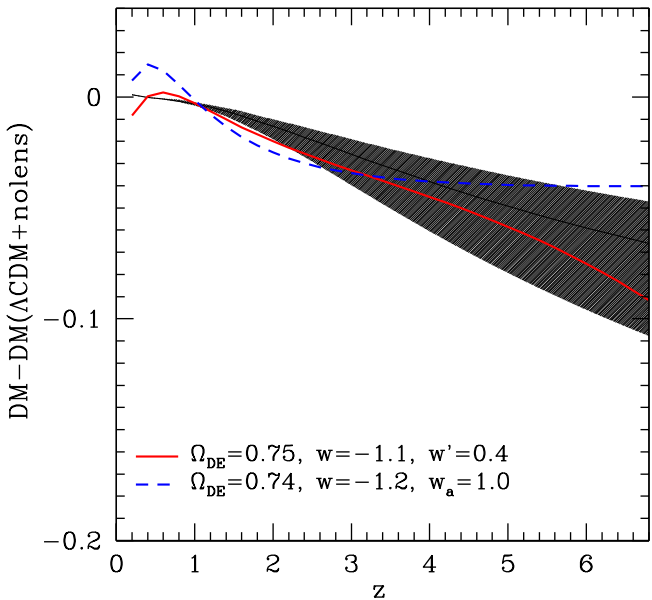}
\caption{The mean deviation of the distance modulus (DM) for {\it Swift}
 GRBs introduced by gravitational lensing, $-2.5\log\langle\mu\rangle$,
 is shown by a thin solid line. The shaded region shows how sensitive
 the magnification bias is to the slope of the luminosity function:
 The width corresponds to the change of the slope by $\pm 0.3$.
  Solid and dashed lines are distance moduli in evolving dark energy
 cosmology models with respect to that in the $\Lambda$CDM model. For
 these lines, the gravitational lensing effect is not considered.} 
\label{fig:dm}
\end{figure}
%%%%%%%%%%%%%%%%%%%%%%%%%%%%%%%%%%%%%%%%%%%%%%%%%%%%%%%%%%%%%%%%%%%%%

Figure \ref{fig:dm} shows how gravitational lensing biases the Hubble
diagram from GRBs. First of all, the degree of the bias again depends
strongly on the redshift. At $z=5$ lensing decreases the distance
modulus by $\sim -0.05$, thus it cannot be ignored. Second, errors of 
slopes of luminosity functions introduce large uncertainties in the
effect of gravitational lensing on the GRB Hubble diagram, in
particular at high redshifts. The shallow luminosity function means
that there are much more GRBs below the flux limit, thus we have much 
more GRBs that are beyond the flux limit just because of amplifications
by gravitational lensing than those which are below the flux limit due
to dimming by lensing, resulting in a large mean magnification.
The dispersion introduced by lensing is $\sim 0.3$ mag at $z\sim 5$,
larger than the systematic shift by the magnification bias. In
summary, we find that gravitational lensing systematically changes the
shape of the GRB Hubble diagram particularly at high redshifts, not to
mention introducing additional dispersions. 

How does the bias affect the determination of the dark energy equation
of state from the GRB Hubble diagram? To explore this, we consider the
following two parameterizations of the equation of state $w(z)$: One is
$w(z)=w_0+[z/(1+z)]w_a$ and the other is  $w(z)=w_0+w'z$, both of
which have been widely adopted in dark energy studies. We note that
in the latter parametrization $w(z)$ increases rapidly with increasing
redshift, which results in unphysical values of $w(z)$ at high
redshifts. We compute distance moduli for following two 
parameter sets, ($\Omega_{\rm DE}$, $w_0$, $w'$)$=$($0.75$, $-1.1$,
$0.4$) and ($\Omega_{\rm DE}$, $w_0$, $w_a$)$=$($0.74$, $-1.2$,
$1.0$), which are plotted in Figure \ref{fig:dm}. As shown in the
Figure, distance moduli of these evolving dark energy models behave
just as that of the $\Lambda$CDM model biased by gravitational
lensing. The increase of the equation of state as a function of $z$
results in the enhancement of the expansion rate at high-redshifts and
thus make distant objects brighter than those in no dark energy
evolution model, which is just similar as gravitational lensing
effect. It is interesting to note that the behavior more resembles
dark energy parametrized unphysically as $w(z)=w_0+w'z$. 
Therefore the effect of gravitational lensing is expected to
appear as an artificial evolving dark energy in determining
cosmological parameters from the GRB Hubble diagram. 

%%%%%%%%%%%%%%%%%%%%%%%%%%%%%%%%%%%%%%%%%%%%%%%%%%%%%%%%%%%%%%%%%%%%%
\begin{figure}[!t]
\includegraphics[width=0.40\textwidth]{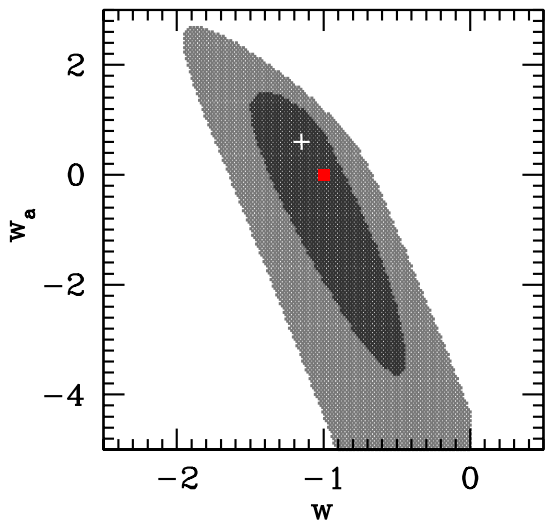}
\caption{Projected constraints on the dark energy equation of state
 expected from 400 {\it Swift} GRBs. Light and dark shaded regions
 indicate 68\% limits for the cases with luminosity inference
 uncertainties of $\sigma=0.6$ mag and $0.16$ mag, respectively. The
 best-fit values for $\sigma=0.6$ is shown by crosses. Filled square
 indicates the $\Lambda$CDM model that we originally assumed in
 simulations of GRBs. For the parameterization of the equation
 of sate, we adopt $w(z)=w_0+[z/(1+z)]w_a$.}   
\label{fig:chi2_w}
\end{figure}
%%%%%%%%%%%%%%%%%%%%%%%%%%%%%%%%%%%%%%%%%%%%%%%%%%%%%%%%%%%%%%%%%%%%%

%%%%%%%%%%%%%%%%%%%%%%%%%%%%%%%%%%%%%%%%%%%%%%%%%%%%%%%%%%%
%%%%%%%%%%%%%%%%%%%%%%%%%%%%%%%%%%%%%%%%%%%%%%%%%%%%%%%%%%%
\section{Monte-Carlo Simulations}
%%%%%%%%%%%%%%%%%%%%%%%%%%%%%%%%%%%%%%%%%%%%%%%%%%%%%%%%%%%
%%%%%%%%%%%%%%%%%%%%%%%%%%%%%%%%%%%%%%%%%%%%%%%%%%%%%%%%%%%

We pursue the bias introduced by gravitational lensing further by
performing Monte-Carlo simulations of GRB observations. Again, we
adopt a model of Firmami et al. \cite{Firmani:2004fn} for the
cosmological distributions of GRBs, and also assume the $\Lambda$CDM
model as the underlying cosmology.  First we distribute GRBs in
redshift-luminosity space according to the GRB formation rates and the
best-fit luminosity function. We are conservative to restrict our
attention to GRBs at $z<7$, because it is still unclear whether we can
really measure redshifts of GRBs at $z>7$. For each event, we first 
compute an observed  flux. After adding a measurement error which is
modeled by a Gaussian error with $\sigma=0.05$ mag, we compare it with
the flux limit of {\it Swift} satellite $P_{\rm lim}=0.4$ ${\rm
  cm^{-2}s^{-1}}$ and select it if the flux is above the limit. In
this way we construct simulated data of 400 GRBs; the number is based
on the fact that {\it Swift} satellite is discovering $\sim 100$ GRBs
per year. We note that the number is somewhat optimistic since only of
order half of {\it Swift} GRBs have measured redshifts.
Next, for each event, we infer its absolute luminosity using
any correlations with observables such as variability and
spectral-lag. Here we do not ask which indicators are used, but just
assume a Gaussian with $\sigma=0.6$ mag as an error associated with
the estimate of the absolute luminosity. This error is
conservative in the sense that one can reduce the error to $\sim 0.45$
mag by making use of several indicators (see \cite{Liang:2005xb}),
but could be optimistic in the sense that it ignores several
systematic effects \cite{Friedman:2004mg}. It
is possible that we find better correlations which significantly
reduce the error in estimating the absolute luminosity. Thus, we also
consider the case $\sigma=0.16$ mag which is comparable to that of
SNeIa. From observed fluxes and inferred absolute luminosities, we can
derive distance moduli for these GRBs, which are used to constrain
cosmological parameters. We compute $\chi^2$ in four parameter space,
$\Omega_{\rm M}=1-\Omega_{\rm DE}$, $h$, $w_0$, and  $w_a$ assuming a
flat universe. We add Gaussian priors $\Omega_{\rm 
  M}=0.27\pm0.04$ and $h=0.72\pm0.04$, which are reasonable given the
current accuracy of cosmological parameter determinations
\cite{Spergel:2006hy}. To reduce offsets of best-fit values by
statistical fluctuations, we repeat the calculation of $\chi^2$ for 30
different realizations of 400 GRBs and take an average of $\chi^2$. 

We show constraints on the equation of state in Figure \ref{fig:chi2_w}.
As expected, gravitational lensing shifts the best-fit models: The
best fit parameter set from fitting is ($w_0$, $w_a$)=($-1.15$,
$0.6$). However, the assumed model is still within 68\% confidence
limit for both large and small errors in estimating GRB absolute
magnitudes, implying that gravitational lensing does not bias the
parameter estimate very much for {\it Swift} GRBs. The cosmological
constant model ($w(z)=-1$ has $\Delta\chi^2\sim 0.5$ when the error is
large and $\Delta\chi^2\sim 0.1$ when the error is small. We note that
the effect of lensing is dependent on the shape of the luminosity
function: Thus the effect could be much more significant if the
luminosity function has a steep slope at high redshifts, as is clear
from Figure \ref{fig:dm}. In summary, the gravitational lensing is
important systematic effect in constraining the nature of dark energy
from the GRB Hubble diagram but is probably not so drastic as to
totally change results from {\it Swift} GRBs. 

%%%%%%%%%%%%%%%%%%%%%%%%%%%%%%%%%%%%%%%%%%%%%%%%%%%%%%%%%%%
%%%%%%%%%%%%%%%%%%%%%%%%%%%%%%%%%%%%%%%%%%%%%%%%%%%%%%%%%%%
\section{Conclusion}
%%%%%%%%%%%%%%%%%%%%%%%%%%%%%%%%%%%%%%%%%%%%%%%%%%%%%%%%%%%
%%%%%%%%%%%%%%%%%%%%%%%%%%%%%%%%%%%%%%%%%%%%%%%%%%%%%%%%%%%

In this paper, we have studied the effect of gravitational lensing on
the determination of dark energy properties from distant GRBs.
Gravitational lensing systematically brighten the apparent luminosity
of observed GRBs, resulting in the modification of the Hubble diagram
inferred from GRBs. Our Monte-Carlo simulations have shown that
lensing does bias the best-fit model, but we can recover the assumed
model within 68\% confidence limit for {\it Swift} GRBs. Therefore, it
does not degrade the cosmological usage of GRBs very much, at least
within accuracies currently expected from {\it Swift} observations.
One of the reasons is that the constraint on dark energy from {\it
 Swift} GRBs are not very strong. We, however, emphasize that the
amount of the bias introduced by lensing is quite sensitive to the
shape of the luminosity function. Thus, we need an accurate
measurement of the GRB luminosity function, in particular near the
flux limit and beyond, in order to remove the effect of gravitational
lensing and to obtain unbiased Hubble diagram. In addition, the strong
influence of gravitational lensing on high-redshift GRBs implies that
estimates of high-redshift GRB (or similarly star formation) rates and
even the luminosity function itself may be biased. We plan to address
these issues in a forthcoming paper.   

%%%%%%%%%%%%%%%%%%%%%%%%%%%%%%%%%%%%%%%%%%%%%%%%%%%%%%%%%%%%%%%%%%%%%
\begin{acknowledgments}
We thank Masahiro Takada, Sudeep Das, Eran Ofek, Vladimir Avila-Reese
for useful discussions. We also thank anonymous referees for useful
suggestions. K. T. is supported by Grant-in-Aid for JSPS Fellows.
\end{acknowledgments}
%%%%%%%%%%%%%%%%%%%%%%%%%%%%%%%%%%%%%%%%%%%%%%%%%%%%%%%%%%%%%%%%%%%%%

\end{document}